# High-space-bandwidth product characterization of metalenses with Fourier ptychographic microscopy

Chuanjian Zheng, [+ab] Wenli Wang, [+ab] Yanfang Ji, [c] Yao Hu, [*ab] Shaohui Zhang, [*ab] and Qun Hao[*ab]

[a] School of Optics and Photonics, Beijing Institute of Technology, Beijing 100081, China.

[b] National Key Laboratory on Near-Surface Detection, Beijing, 10012, China

[c] Tianjin H-Chip Technology Group Corporation, Tianjin 300480, China.

[+] These authors contributed equally to this work.

[*] Correspondence: huy08@bit.edu.cn, zhangshaohui@bit.edu.cn, qhao@bit.edu.cn.

**Abstract**

Large numerical aperture (NA) and large aperture metalenses have shown significant performance and abundant applications in biomedical and astronomical imaging fields. However, the high space-bandwidth product (SBP) requirements for measuring the phase of these metalenses, characterized by small phase periods and large apertures, have resulted in no effective techniques for sufficient characterization. In this paper, we propose a high SBP phase characterization technique using Fourier ptychographic microscopy (FPM), enabling a high spatial resolution and wide field of view simultaneously. To demonstrate the feasibility and effectiveness of this technique, we achieve a high SBP (4.91 megapixels) measurement and characterization for focusing and focusing vortex metalenses, quantitatively displaying the effect of fabrication error on their typical optical performance. Furthermore, we characterize the aberration type and amount of wavefront deviations caused by fabrication. We also analyze compensation methods for different aberrations based on the wavefront characterization results, providing a targeted alignment strategy for optimizing overall optical system performance. We believe that our high SBP characterization technique cannot only help to improve metalens design but also optimize its fabrication processing, which will pave the way for the diversified applications of metalenses.

**Introduction**

Due to the great development of design and nanofabrication technologies [1–4], the large numerical aperture (NA) metalens [5] can be easily realized and the aperture of the metalens has also been able to reach the centimeter scale [6]. These breakthroughs make engineering optics into a new era: Engineering Optics 2.0 [7]. Currently, metalenses can not only be applied to biomedical imaging [8], like high-resolution imaging [9], quantitative phase imaging [10], but also bring new development opportunities to the field of large-aperture optics [6,11], such as astronomical imaging [12] and remote underwater surveillance [13], as well as in high-power lasers [14]. These potential applications are all inseparable from precise phase manipulation of metalenses. Although the designed phase distributions of metalenses can usually be perfectly obtained from numerical simulations, the real phase distributions generated by metalenses are often limited by materials as well as nano-manufacturing techniques [15–17]. These limitations usually cause phase deviation that cannot be ignored, which usually changes the optical performance of metalens-based devices [16]. Therefore, a reliable optical phase measurement and characterization technique is urgent to facilitate comparisons between designed and actual metalenses.

So far, near-field scanning [18,19] and interferometry [16,20–22] are two most common ways for quantitative phase measurements of metalenses. Near-field optical scanning can offer sub-diffraction resolution but is pretty time-consuming and invasive [18,19]. Thus, this technique has never been reported for the whole aperture metalens phase measuring. Several interferometric methods, have been reported to measure the phase of metalenses, such as Mach-Zender interferometry [16], quadriwave lateral shearing interferometry [21], etc. However, the dispersion compensation, the precise delay, the high mechanical stability, etc., required may limit their application environment in the optical phase measurement and characterization of metalenses. More importantly, the interferometric methods are limited by the space-bandwidth product (SBP) of the optical systems, which equals the number of optically resolvable spots within a field of view [23], resulting in a trade-off between the spatial resolution and field of view. However, the high SBP is indispensable for measuring large NA and large aperture optical metalenses. Specifically, on the one hand, the large NA metalens contains adjacent units with a large phase gradient, which indicates that the number of unit structures required in one phase period is relatively small, leading to its high spatial frequency. So, high resolving power is necessary. On the other hand, the phase measurement for large aperture metalens is also inseparable from the large field of view, all of which means interferometric methods are difficult to measure.

In addition, the transport-of-intensity-equation-based method, as a classical non-interferometry-based technique, has also been effectively demonstrated for wavefront aberration calculation of metalenses[24,25]. This method is much less restrictive in terms of experiment requirements due to strong anti-interference performance [3,26]. However, phase measurement based on the transport-of-intensity equation depends on multiple mechanical movements and two restrictive assumptions: weak defocusing and paraxial approximation [24,27–29], which can cause low phase measurement accuracy and limited SBP, hindering the phase measurements of large NA and large aperture metalenses. To the best of our knowledge, metalens phase measurement techniques that can simultaneously satisfy high accuracy and high SBP have not been reported. In view of the measuring and characterizing requirements for large NA and large aperture metalenses, exploring a simple, high accuracy, and high SBP technique for metalens optical phase measurement and characterization is extremely urgent.

In this article, we propose a high SBP phase characterization technique with Fourier ptychographic microscopy (FPM) to meet the requirements for phase measurement and characterization of large NA and large aperture metalens. The technique achieves high accuracy (0.02 rad) and high SBP (4.91 megapixels) phase measurement while offering strong anti-interference performance without dispersion compensation and precise delay, as well as loads of mechanical movements. It brings great support for phase measurement and characterization of large NA and large aperture metalens. Based on this technique, we measured a focusing metalens and a focusing vortex metalens in experiments since they both have significant applications in biomedical imaging and astronomical imaging. With the measured phase distribution, we quantitatively characterized the optical properties of the metalenses, including focal length, point spread function (PSF), depth of focus (DOF), Strehl ratio (SR), modulation transfer function (MTF) and topological charge (TC), etc. These optical parameters can effectively help to illustrate the effect of fabrication error on their work performance. Moreover, we also analyzed the type and amount of wavefront deviations between the designed and measured values of metalenses using Zernike polynomials, which can assist the alignment of the metalenses in optical systems and improve the design and fabrication processing.

**Results**

*Phase measurement based on the FPM*

The essential problem in most phase measurement techniques is recovering the missing phase distribution from the captured intensity images. To obtain the metalens phase information, transforming the phase information to recordable intensity variation by encoding the input optical field and recovering the phase information by decoding intensity information, i.e., encoded record and decoded recovery, are necessary. For example, phase shift encoded and propagation encoded are usually adopted in interferometry and TIE techniques, respectively. However, the SBP of these methods is limited to the coherent spectral bandwidth, as shown in the inset figure of Fig. 1, which hinders their applications for measuring large NA and large aperture metalenses.

In our proposed method, a new encoded record strategy, angle-varied illumination, is proposed. As shown in Fig. 1, multiple angle-varied plane beams are used to illuminate the metalens sequentially, and resultant intensity images are captured as the raw data. As the illumination angle increases, not only is the phase information encoded as intensity variations in captured images, but the spectral bandwidth of the phase information is also expanded. Thus, a high SBP phase distribution can be easily decoded without the necessity of extra hardware and complex stitching methods. Moreover, our technique benefits from the much less restrictive experiment requirements of non-interferometric methods and avoids extensive mechanical scanning while maintaining high accuracy. It enables a simple, high-accuracy and high SBP phase measurement. It brings great hope to phase measurement and characterization of large NA and large aperture metalenses. The details of the FPM experimental setup can be seen in Supplementary Information SI.1. Since the high-accuracy phase measurement with conventional FPM is still a challenging task due to imperfect capturing and post-processing strategies. We also carefully optimized the working processing for accurate metalens phase measurement, which can be seen in Supplementary Information SI.2 and SI.3.

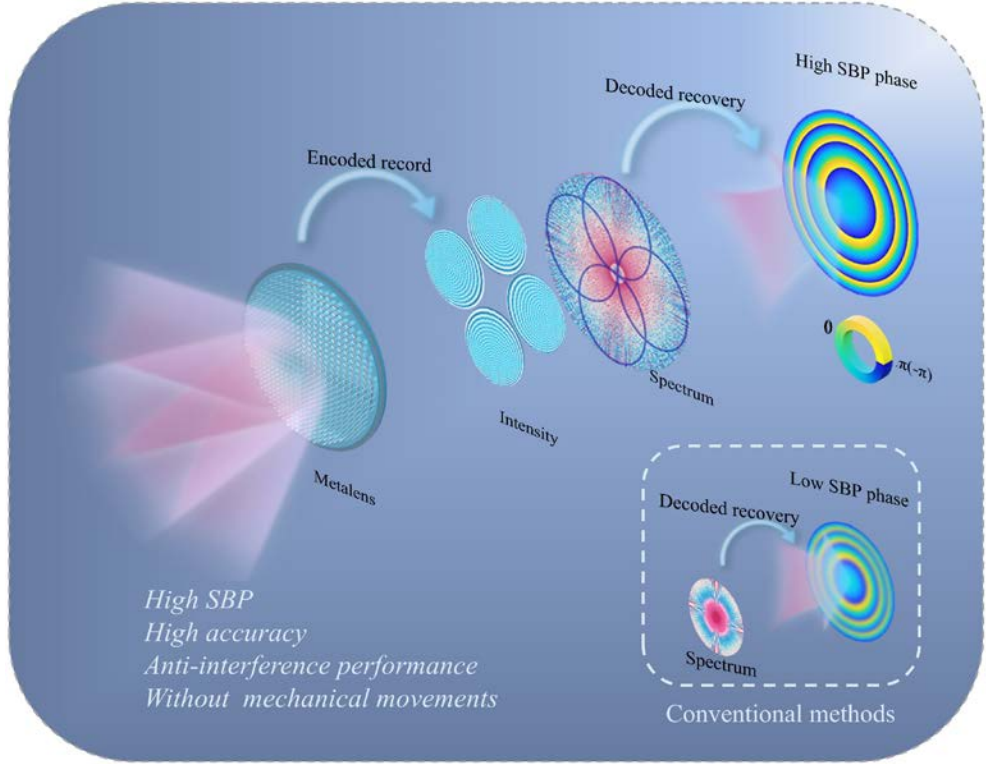

Fig. 1. Illustration of the high SBP phase measurement technique with Fourier ptychographic microscopy. The inset figure shows the scheme of phase measurement based on conventional methods.

## SBP and phase measurement accuracy of FPM experimental setup

To verify the SBP enhancement capability and phase measurement accuracy of the FPM experimental setup, we measured the phase distribution of a quantitative phase target (Benchmark Technologies). The phase target we chose consists of multiple structures with the same height (50 nm) but different line patterns. We captured 225 low-resolution (LR) intensity images under angle-varied illumination and fused them according to the FPM procedures (see details in Supplementary Information SI.2). Fig. 2(a) shows the captured LR intensity image under normal illumination. Fig. 2(b) displays the measured phase distribution. We also measured these structures using a standard white light interferometer for reference, as shown in Fig. 2(c).

    To clarify the enhancement capability of the technique on the SBP, we analyzed the results in Figs. 2(a) and 2(b). The line patterns a1 and a2 within the white dashed boxes in Fig. 2(a), respectively, represent the intensity images of the minimum structures that can be resolved by the system before and after the method is used. Similarly, the line patterns b1 and b2 within the orange dashed boxes in Fig. 2(b), respectively, represent the measured phase distribution of the corresponding minimum structures. The line pattern a1 confirms the original full-pitch lateral resolution as 1.38 μm in Group 9, Element 4, which agrees well with the theoretical coherent

diffraction-limit lateral resolution ($\lambda/NA_{obj}$ =1.25 μm, where $\lambda$ = 0.623 μm is the wavelength, $NA_{obj}$ denotes the numerical aperture of the objective). Meanwhile, the line pattern b2 in Group 10, Element 4, with the whole of Group 10 illustrated as a magnified phase pattern in Fig. 2(d) and cross-sectional line in Fig. 2(e) for better visualization, shows an improved full-pitch lateral resolution of 0.69 μm. Given the field of view of ~0.58 mm$^2$, we can obtain an increase of SBP from 1.23 megapixels to 4.91 megapixels.

Moreover, to describe the phase accuracy, we picked line patterns b1 and b2 within the dashed boxes in Fig. 2(b) and compared their phase distributions with those of the line patterns c1 and c2 in the dashed boxes in Fig. 2(c). By comparing the line patterns b1 and c1, respectively, illustrated as magnified phase patterns in Figs. 2(b1) and 2(c1), we can obtain the RMS value of the phase error is ~0.02 rad, as shown in Fig. 2(f1). A similar conclusion can be obtained when comparing the minimal resolvable line patterns b2 and c2, respectively, illustrated as magnified phase patterns in Figs. 2(b2) and 2(c2), as shown in Fig. 2(f2). Thus, the phase accuracy of our method is ~0.02 rad.

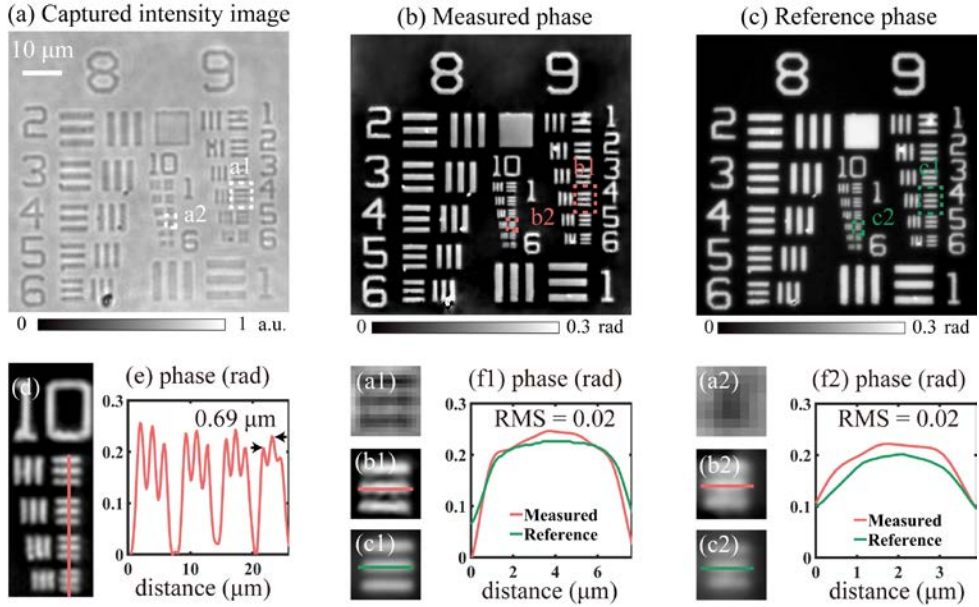

Fig. 2. Phase measurement results of the quantitative phase target. (a) captured LR intensity image of the 50 nm-height structure. Line patterns a1 and a2 within dashed boxes, respectively, represent the intensity image of the minimum structures that can be resolved by the system before and after the method is used. (b) measured phase distribution of the 50 nm-height structure. Line patterns b1 and b2 within dashed boxes, respectively, represent the measured phase distribution of the minimum structures that can be resolved by the system before and after the method is used. (c) reference phase of the 50 nm-height structure. Line patterns c1 and c2 within dashed boxes, respectively, represent the reference phase of the minimum structures that can be resolved by the system before and after the method is used. (d)

measured phase distribution of Group 10 in 2(b). (e) measured phase line profile along the line pattern in 2(d). (a1), (b1), and (c1) respectively represent the enlarged line patterns a1, b1, and c1 within dashed boxes. (f1) phase line profiles of line patterns in 2 (b1) and 2(c1). (a2), (b2), and (c2) respectively represent the line patterns a2, b2, and c2 within dashed boxes. (f2) phase line profiles of line patterns in 2 (b2) and 2(c2).

In general, our method can achieve high phase accuracy and high SBP. The phase accuracy of our FPM system is ~0.02 rad, and the improved SBP is 4.91 megapixels. To our best knowledge, the highest interferometric phase measurement accuracy for metalenses is ~0.05 rad, the full-pitch lateral resolution is 2 μm (0.69 μm in our method), and the SBP is limited by the optical system [16]. Therefore, compared with current interferometric methods, our system offers not only strong anti-interference performance without strict experiment requirements but also offers high accuracy and high SBP without the help of extremely high magnification objectives and complex stitching methods. When compared with other non-interferometry-based techniques, like TIE [24], our system not only breaks through SBP limitations but also avoids multiple mechanical movements in experiments while maintaining high accuracy. To be specific for high SBP, regardless of interferometric methods or TIE method, if one wants to achieve the same amount of SBP as our technique, only can they simultaneously use a higher-magnification objective (60x) and multiple mechanical stitching (3 × 3). It is undoubtedly an extremely high request for the optical system itself and the accuracy of the stitching algorithm. Since the high accuracy, high SBP, strong anti-interference performance and simple procedures of our method, it is obvious that our system, based on the FPM technique, is reliable for measuring and characterization of metalens optical phase. It is worth mentioning that the capability to break through SBP limitations also offers indispensable support for metalenses phase measurement with large NA and large aperture.

*Experimental results of measuring and characterizing the metalenses*

According to the above experiments, we have demonstrated that our method has a great potential to measure and characterize the metalenses. Since the focusing metalens and focusing vortex metalens both have significant applications in biomedical imaging and astronomical imaging and have urgent requirements for large NA and large aperture, we now focus on these two most common metalenses. The relevant details of their design can be seen in Section **Metalenses design and simulation**. Their designed amplitude and phase distributions are shown in Supplementary Information SI.4.

The measuring and characterizing results of the focusing metalens based on the FPM method are displayed in Fig. 3. This focusing metalens can converge a plane beam into a spot at its focal plane, as shown in Fig. 3(a). Fig. 3(b) displays the captured LR intensity image under normal illumination. The measured wrapped phase distribution is shown in Fig. 3(c). Since the unwrapped phase distribution makes more sense than the wrapped phase distribution; we directly compared the measured and designed unwrapped phase distributions [ Figs. 3(d1) and 3(d2)], which indicate that they resemble although the deviation is existent. Quantitative phase deviation distribution is also calculated and shown in Fig. 3(e), where the RMS and PV values are 2.507 rad and 16.168 rad, respectively. We can also find that most of the large phase deviation exists in the edge area of the metalens. We plotted the data positioned along the red dotted line in Figs. 3(d1) and the green dotted line in Figs. 3(d2), in Fig. 3(f) for clearer visualization, which shows the two curves are basically matched and have obvious convergence characteristics except for the phase information at the edge area (x-coordinate over 21 μm, as denoted by the red arrow). Given that the minimum phase period (1.13 μm) of the wrapped phase distribution of the focusing metalens is larger than the lateral resolution of our system (0.69 μm) and the system error has been removed (see details in Supplementary Information SI.2), we think the FPM system can correctly evaluate the phase distribution within the whole aperture of the focusing metalens. The phase deviation should be caused by the fabrication error, and the deviation at the edge area is more significant than that at the central area.

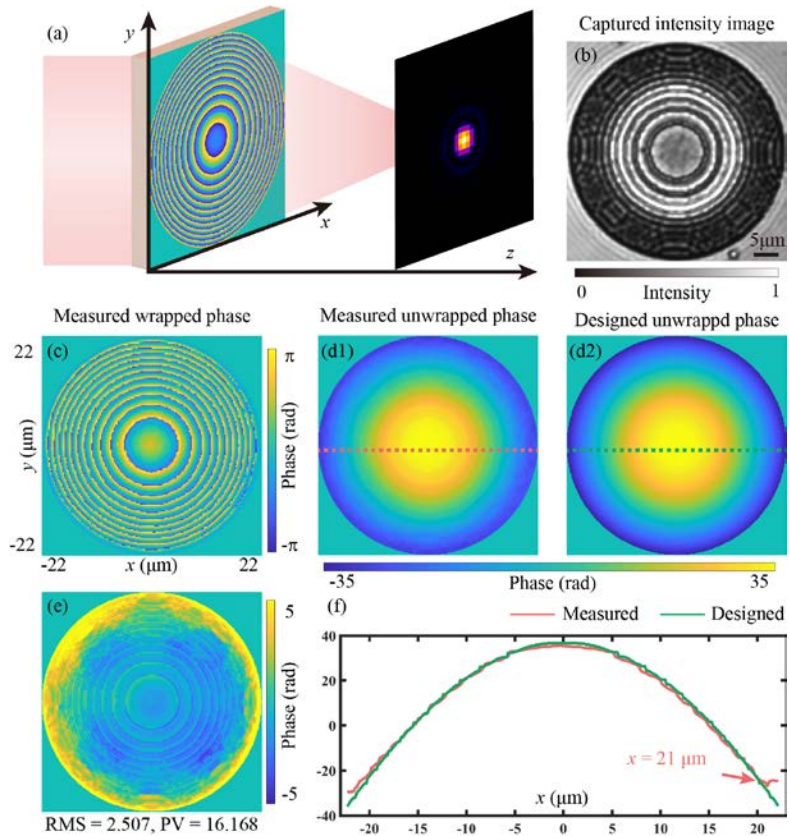

Fig. 3 Experimental measuring results of the focusing metalens based on the FPM method. (a) a function diagram of the focusing metalens. (b) captured intensity image under normal illumination. (c) measured wrapped phase distribution. (d1) measured unwrapped phase distribution. (d2) designed unwrapped phase distribution. (e) measured

phase deviation. (f) phase profiles, the red line represents the measured phase distribution in (d1), and the green line represents the designed phase distribution in (d2).

We also experimentally measured and characterized the focusing vortex metalens, as displayed in Fig. 4. This focusing vortex metalens can converge a plane beam into an annular intensity distribution at its focal plane, as shown in Fig. 4(a). Figs. 4(b) and 4(c), respectively, show the captured LR intensity image under normal illumination and the measured wrapped phase distribution. By comparing the measured and designed unwrapped phase distributions [Figs. 4(d1) and 4(d2)], we can find their patterns are much similar. Their quantitative deviation is shown in Fig. 4(e), where the RMS and PV values are 0.555 rad and 6.739 rad, respectively. To further visualize their deviation, we displayed the data positioned along the red dotted line in Figs. 4(d1) and the green dotted line in Figs. 4(d2), as shown in Figs. 4(f), from which we can see that the designed and measured phase curves are basically matched, but the lateral dimensions of the curves are not consistent. It means that the focusing vortex metalens we fabricated has an obvious lateral dimensional negative error. Since the effect of system aberration has been removed, these deviations should be all fabrication errors.

Based on the experimental results of the two metalenses, we believe that our FPM technique can measure and characterize metalenses effectively in a simple way. Moreover, in view of the high SBP of our method, this technique should also be able to directly measure and characterize the metalenses with large NA and large aperture.

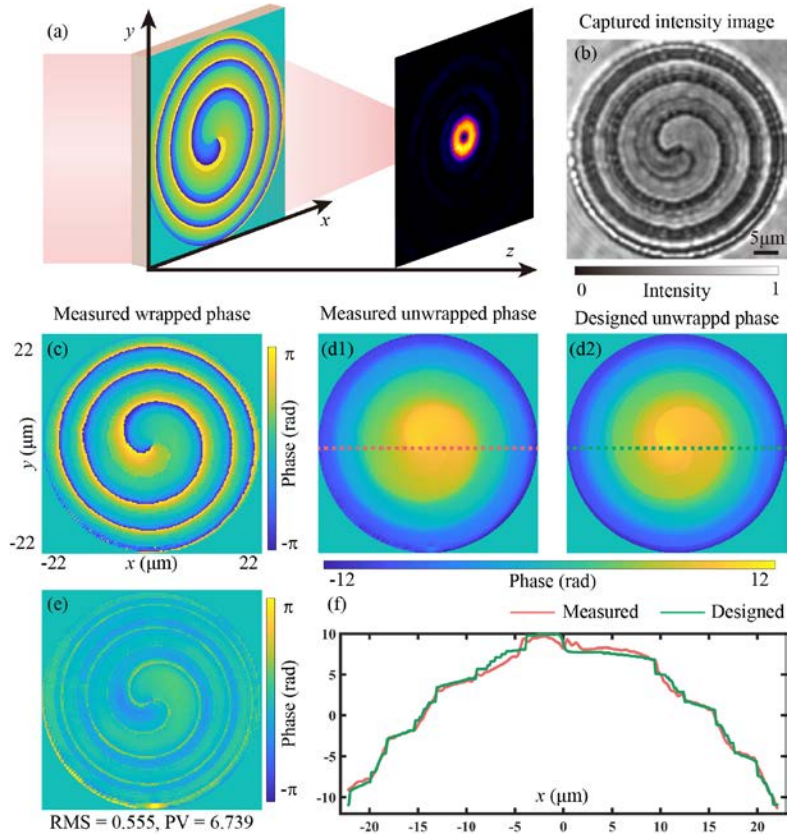

Fig. 4. Experimental measuring results of the focusing vortex metalens based on the FPM method. (a) a function diagram of the focusing vortex metalens. (b) captured intensity image under normal illumination. (c) measured wrapped phase distribution. (d1) measured unwrapped phase distribution. (d2) designed unwrapped phase distribution. (e) measured phase deviation. (f) phase profiles, the red line represents the measured phase distribution in (d1), and the green line represents the designed phase distribution in (d2).

*Analysis of optical performance*

To further characterize the optical performance of these two metalenses, we used the phase distributions to calculate relevant optical parameters and evaluate the effect of phase deviation on these parameters.

For the focusing metalens, the focal length is an indispensable indication for evaluating the ability of the metalens-based optical system to converge light. It plays a decisive role in the position of the image, the scale of the object and the image, the size of the field of view and the size of the depth of field. To calculate its focal length, we first used the angular method [30] to calculate the light intensity distributions on the vertical-section plane ($x$-$z$). As shown in Figs. 5(a1) and 5(a2), respectively, show the intensity distributions calculated according to their designed and measured phase distributions. From the inset figures in those two figures, the calculated focal lengths are displayed, respectively, 31.6 μm and 30.6 μm. It is obvious that the measured focal length is a little smaller, and there is a 1 μm focal length deviation between them, which the phase deviations between the designed and measured values should cause.

In addition, PSF can be used to evaluate the ability of optical systems to retain detail or analyze the aberration of the optical system. DOF can be used to assess the range for clear imaging on both sides of the focal plane. SR can be used to quantify the ability of optical systems to focus. MTF can be used to evaluate the imaging performance of optical systems at different frequencies, including contrast, sharpness, and color reproduction. Thus, for the focusing metalens, we also calculated the above optical parameters to evaluate its optical imaging performance more comprehensively and provide guidance for its specific application in the field of biomedical imaging or astronomical imaging. All of these optical parameters are calculated according to the methods in Supplementary Information SI.5, and their values are displayed in Fig. 5.

Figs. 5(b1) and 5(b2) show the PSF distributions of the focusing metalens calculated according to the designed and measured phase distributions, respectively, where the inset figures both show the logarithm of the PSF. We can find that the peak value of PSF in Fig. 5(b1) is larger than that in Fig. 5(b2), which means that the imaging equality of this measured focusing metalens is worse than that of the designed one due to the fabrication error. It can also be seen that there is a slightly asymmetric distribution in Fig. 5(b2), which means that the fabrication error may generate asymmetric aberration.

The DOF, which is equal to the full width at 80% maximum of the PSF [16], is shown in Fig. 5(c). The DOF value calculated according to the designed phase distribution and the measured phase distribution are, respectively, 1.7 μm and 1.8 μm. This result indicates that when using the focusing metalens to obtain a satisfactorily clear image of the object, the accepted distance of the object can move forward and backward is about 1.8 μm.

To calculate the SR = $I_{max\_Measured}$/ $I_{max\_Designed}$, i.e., the ratio of the peak value of PSF in Fig. 5(b2) and Fig. 5(b1), the PSF curve was also plotted in Fig. 5(d), showing the SR values is 0.62. This SR value indicates that the fabrication error of the metalens severely reduces the light intensity of the focus.

The MTF is the amplitude of the Fourier transform of PSF. We displayed the corresponding MTF curves of the focusing metalens in Fig. 5(e), from which we can see the measured curve exhibits severe contrast loss because of the fabrication error, especially at low frequency. Therefore, the fabricated focusing metalens does not have a good object contour transfer capability.

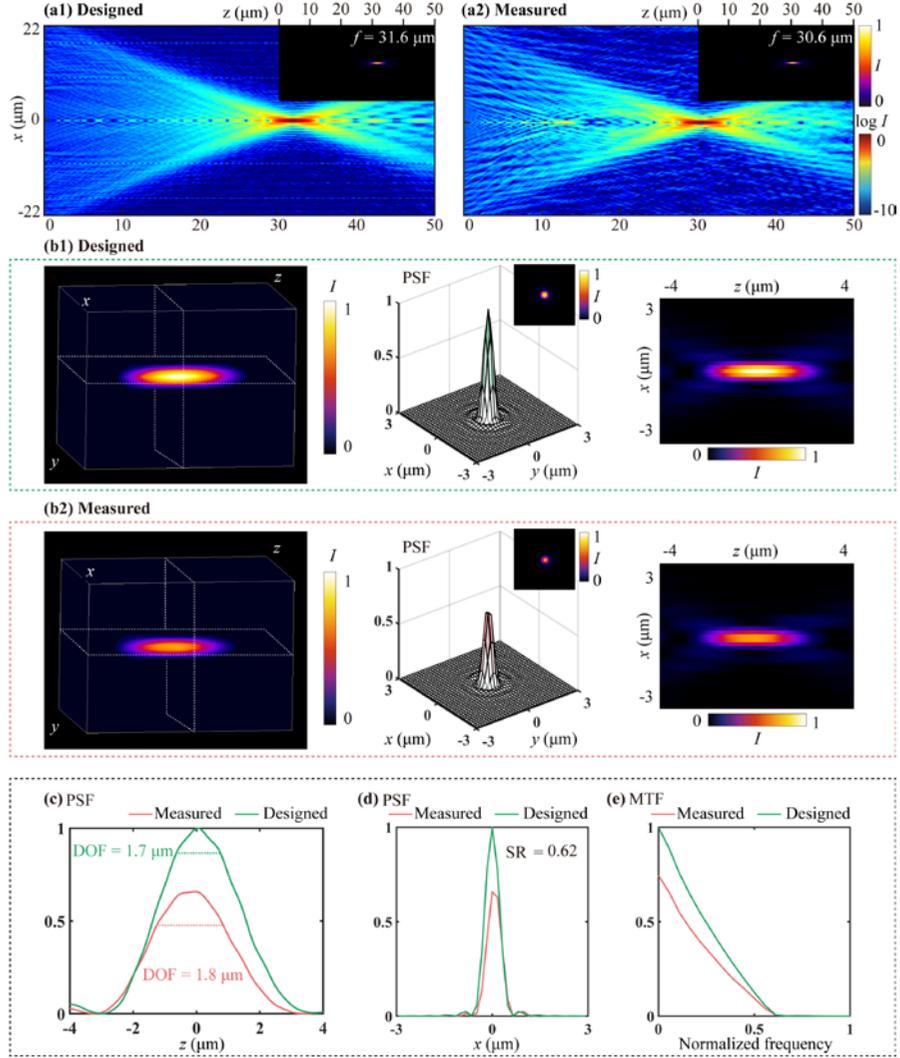

Fig. 5. Optical performance characterization of the focusing metalens. (a1), (a2) calculated light intensity distributions on the x-z plane and focal length according to the designed and measured phase distributions. (b1) calculated PSF according to the designed phase distribution, and the inset shows the logarithm of the PSF. (b2) calculated PSF according to the measured phase distribution, and the inset shows the logarithm of the PSF. (c) DOF values. (d) SR value (here, SR=$I_{max}$_Measured/ $I_{max}$_Designed). (e) MTF curves.

For the focusing vortex metalens, the focal length is also a vital optical parameter as that for the focusing metalens. The TC value is another important parameter that can help measure the orbital angular momentum of a vortex. It is of important research significance in many fields, such as optical imaging, measurement, and communication.

To obtain the focal length of the focusing vortex metalens, we used the same method as that of the focusing metalens. Figs. 6(a1) and 6(a2), respectively, show the intensity distributions on the *x-z* plane of the focusing metalens according to the designed and measured phase distributions, and the calculated focal lengths are, respectively, 122.9 μm and 122.7 μm. We can find that there is a 0.2 μm focal length deviation between them due to the phase deviations between the designed and measured distributions.

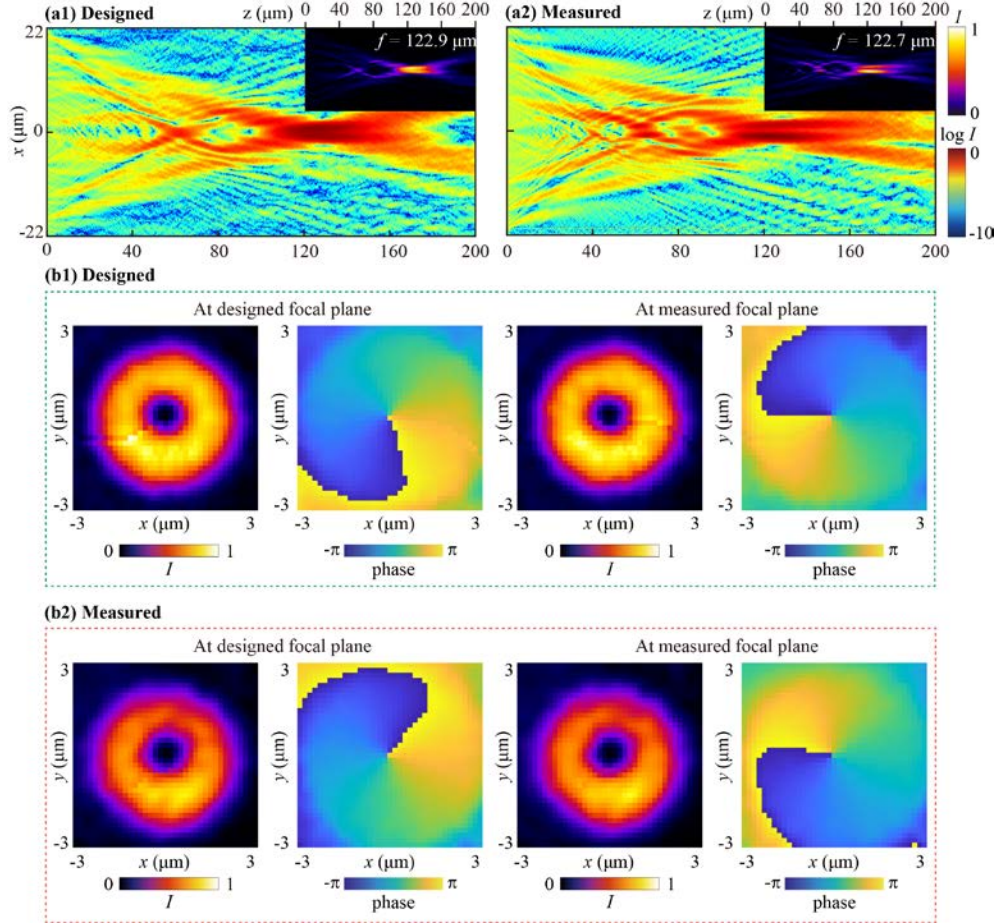

Fig. 6. The optical performances of the focusing vortex metalens. (a1), (a2) calculated light intensity distribution on the *x-z* plane and focal length according to the designed and measured phase distributions. (b1) calculated light field distributions at different focal planes according to the designed phase distribution. (b2) the calculated light field distributions at different focal planes according to the measured phase distribution.

To obtain the TC values, we calculated the intensity and phase distributions at both designed and measured focal planes by using the designed and measured phase distributions, respectively. Fig. 6(b1) shows two obvious toroidal intensity distributions and spiral phase distributions calculated according to the designed phase distribution, indicating the generated beams are both vortex beams with a TC equal to 1. A similar conclusion can be obtained when the light field is calculated according to the measured phase distribution, as displayed in Fig. 6(b2). When comparing the results in Figs. 6(b1) and 6(b2), we can see their TC values are equal no matter which focal plane we observed, although their intensity and phase distributions are various. So, for the focusing vortex metalens, we can hold on to the fact that the phase deviation between the designed and the measured phase distribution does not affect TC values.

From the above calculations and analyses of optical performance, the relevant evaluation index difference caused by phase deviation can be clearly obtained, which not only can effectively evaluate the work performance of metalenses but also illustrates the effect of fabrication error on their working performance. This work may more clearly guide the techniques advanced in metalens optimization design and fabrication to improve its optical performance further.

*Analysis of wavefront deviation*

To further identify and compensate for the deviation between measured and designed phase distributions and assist the alignment of the metalenses in optical systems, we also used Zernike coefficients to characterize the wavefront deviations of the two metalenses. The wavefront deviation can be expanded in terms of a complete set of Zernike circle polynomials that are orthogonal over a unit circle with $r \in [0,1], \phi \in [0, 2\pi]$ in the form of [31]

$$W(r,\phi) = \sum_j C_j Z_j(r,\phi), \tag{1}$$

where $C_j$ are expansion coefficients. The Zernike polynomials $Z_j$ are defined as [32]
when $b$ is even and $b \neq 0$,

$$Z_j(r,\phi) = \sqrt{2(a+1)} R_a^b(r) \cos(b\phi), \tag{2}$$

when $b$ is odd and $b \neq 0$,

$$Z_j(r,\phi) = \sqrt{2(a+1)} R_a^b(r) \sin(b\phi), \tag{3}$$

when $b=0$,

$$Z_j(r,\phi) = \sqrt{(a+1)} R_a^0(r), \tag{4}$$

where $a$ and $b$ are positive integers, including zero, following the restrictions $a - b \geq 0$, and even. The index $j$ is a polynomial-ordering number and is a function of $a$ and $b$. $R_a^b(r)$ are defined as

$$R_a^b(r) = \sum_{s=0}^{(a-b)/2} \frac{(-1)^s r^{a-2s}(a-s)!}{s!(\frac{a+b}{2}-s)!(\frac{a-b}{2}-s)!}. \tag{5}$$

Then, the Zernike coefficients $C_j$ can be retrieved using

$$C_j = \frac{1}{\pi} \int_0^1 \int_0^{2\pi} W(r,\phi) Z_a^b(r,\phi) r dr d\phi. \tag{6}$$

As shown in Fig. 7(a), we used 11 Zernike polynomials to analyze the wavefront deviations, which represent, respectively, as follows: piston, $x$ tilt, $y$ tilt, defocus, primary astigmatism at 45°, primary astigmatism at 0°, primary $x$ coma, primary $y$ coma, trefoil $x$, trefoil $y$ and primary spherical. According to the above equations and the obtained wavefront deviation of focusing metalens [Fig. 7(b1)] and wavefront deviation of focusing vortex metalens [Fig. 7(c1)], we can calculate the corresponding Zernike coefficients up to $j$=11, as shown in Figs. 7(b2) and 7(c2), which obviously display various aberrations and their quantities.

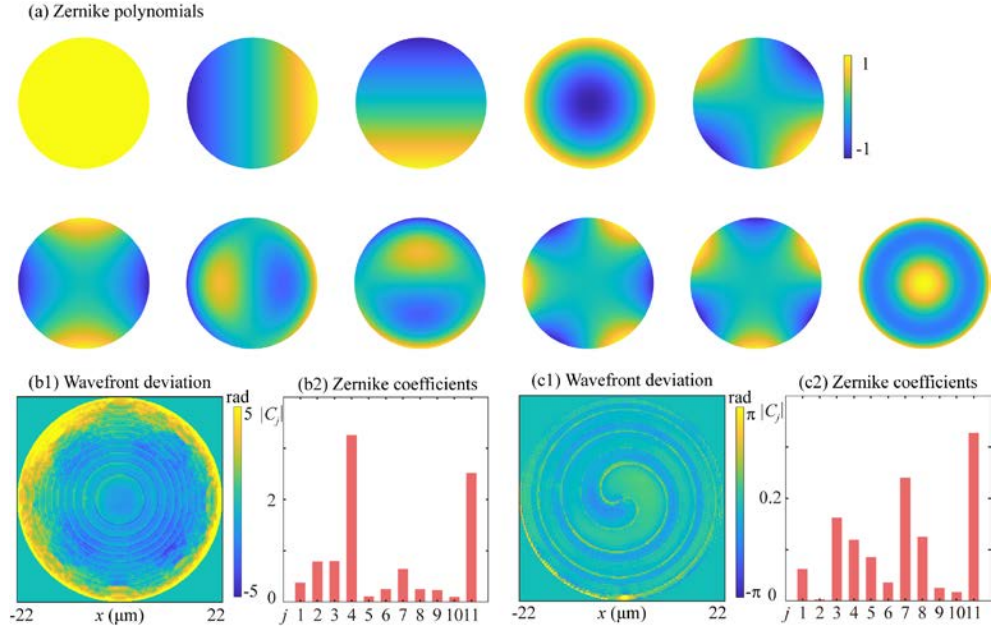

Fig. 7 Analysis of wavefront deviation. (a) real parts of Zernike polynomials up to $j$=11 (b1) wavefront deviation of the focusing metalens; (b2) Zernike coefficients of the corresponding wavefront deviation. (c1) wavefront deviation of the focusing vortex metalens; (c2) Zernike coefficients of the corresponding wavefront deviation.

For the focusing metalens, we find that the defocus and primary spherical aberration account for a large proportion of the overall aberration; the piston, $x$ tilt, $y$ tilt and primary $x$ coma aberrations are the next. However, we usually remove the piston directly in the optical surface analysis; thus, this wavefront deviation mainly causes defocusing, primary spherical aberration, $x$ tilt, $y$ tilt, and primary $x$ coma aberration in an optical system.

It is worth mentioning that the defocusing can be compensated by adjusting the axial distance of the image plane, and the tilt aberrations do not have much effect on the work performance of the optical system, which can be compensated by tilting the metalens. The coma can also be effectively compensated by adjusting the eccentricity and tilting of the metalens. Astigmatism is more troublesome than others mentioned above. Still, the good thing is that it's small, and we can also eliminate it a bit by adjusting the eccentricity and tilt of the metalens. However, spherical aberrations and trefoil are very difficult to eliminate. We often need to optimize the optical design or adjust the experimental optical path, like adding extra optical elements to compensate for these aberrations. In any case, this analysis of wavefront deviation provides a powerful guide for the overall design and installation of optical imaging systems that include focusing metalenses, which can pave the way for the further application of focusing metalenses in biomedical imaging and astronomical imaging.

As for the focusing vortex metalens, it can be found that the piston, *y* tilt, defocus, primary *x* coma and primary spherical aberration are also large, which is the same as that in the focusing metalens. Differently, the primary astigmatism at 45° and primary *y* coma aberrations are still obvious. Since the piston is often removed directly in the optical surface analysis, it can also be concluded that this wavefront deviation of the focusing vortex metalens mainly cause *y* tilt, defocusing, primary coma, spherical aberration as well as astigmatism in an optical system.

In addition, the overall wavefront deviation in the focusing vortex metalens is much smaller than that in focusing metalens. This phenomenon is reasonable because the measured PV and RMS values of the focusing vortex metalens are smaller. These aberrations can also be compensated using the adjustment methods mentioned above. We believe the analysis of wavefront deviation is extremely helpful in identifying and compensating for the performance of these metalens-based devices. It also has great potential to direct the further design and fabrication of metalenses, which will offer powerful help for the diversified applications of metalenses.

**Conclusion**

In conclusion, we developed a high SBP measuring technique based on the FPM to measure and characterize the metalenses. The method can measure the phase distribution of metalenses in a simple, high SBP and high accuracy way, which brings great hopes for characterization of large NA and large aperture metalenses. We demonstrated the feasibility and effectiveness of our technique from the theory and experiments. The phase measurement accuracy is confirmed as ~0.02 rad (RMS), and the systematic SBP increases from 1.23 megapixels to 4.91 megapixels without the necessity of extremely high magnification objectives and complex stitching methods. Two typical metalenses were measured experimentally, and several classical optical parameters were calculated by using the measured and designed phase distributions. The differences in optical parameters caused by the phase deviations between the designed and measured phase distributions were clearly visualized, which can help to quantitatively illustrate the effect of fabrication error on their optical performances. Moreover, Zernike polynomials were used to characterize the wavefront deviations. The type and quantity of aberrations were also analyzed, and the corresponding system adjustment methods were proposed, which will be useful for guiding the overall design and installation of optical imaging systems with metalenses.

**Discussion**

In this article, our technique and optical system do not need a dispersion compensation, a precise delay, high mechanical stability, multiple mechanical scanning, as well as a trade-off between the spatial resolution and field of view, which are often required by common interference techniques or other existing techniques. In particular, we need to emphasize the high SBP feature of our phase measurement technique. Suppose we want to use any other method to achieve the same amount of SBP as our technique, a higher magnification objective (60x) and multiple mechanical stitching (3 × 3) must be used simultaneously. It is undoubtedly a high request for the optical system itself and the accuracy of the stitching algorithm. Therefore, from the perspective of accuracy, SBP, system complexity, and anti-interference performance, our phase measurement technique is most suitable for the measurement and characterization of large NA and large aperture metalenses to the best of our knowledge.

Also, we highlighted not only the benefits of our technique but also its limitations. It requires the capturing and processing of hundreds of images, which results in the decrease of temporal resolution. In order to achieve faster imaging speed, we can replace the LED array with a fiber with higher brightness and then use the galvanometer to achieve the scanning of the lighting angle. In terms of data processing, parallel processing or deep learning can be used to achieve rapid phase information reconstruction.

We can envision our technique becoming a new gold standard in nanophotonics metrology. It will not only assist the alignment and aberration compensation of the metalenses in optical systems but also help to optimize the designs and fabrications of metalenses, which will pave the way for metalenses in future industrial applications, especially for biomedical and astronomical imaging.

**Materials and Methods**

**Metalenses design and simulation**

We designed two metalenses, namely the focusing metalens and the focusing vortex metalens, to demonstrate the feasibility of the FPM technique for measuring and characterizing. These metalenses are composed of a series of nanobricks. Each silicon nitride nanobrick sits on a fused silica substrate for the focusing metalens and the focusing vortex metalens. The transmitted phase of these metalenses depends on the size of nanobricks. These nanobricks are periodically arranged with a fixed square lattice constant $P_x=P_y=500$ nm and a height $H=600$ nm. The simulated propagation phase can basically cover $0\sim2\pi$ by changing the radius $R$ of the nanobrick (See Fig. S7 in Supplements Information SI. 6). The simulated phase $\delta_x$ for a linear polarization incident beam in the $x$-direction (XLP) as a function of $R$ is shown in Fig. S7(a). The corresponding transmission coefficient distribution $t_x$ is exhibited in Fig. S7(b). The wavelength of the simulated incident beam is 623 nm. In this way, we can choose 13 discrete phases generated by nanobricks with different radii, and then a focusing metalens can be constructed by many of these different nanobricks. For a focusing vortex metalens, we changed the number of discrete phases to 20. The designed wrapped phase profiles of these two metalenses are respectively shown in Figs. S6(a) and (c) of Supplements Information SI. 4. The designed amplitude profiles of these two metalenses are respectively shown in Figs. S6(b) and (d) of Supplements Information SI. 4.

**Metalenses fabrication**

Focusing metalens

A 600-nm-thick silicon nitride film is deposited on a quartz substrate by a Plasma Enhance Chemical Vapour Deposition (PECVD) process. Positive resist ZEP520A with a thickness of 170 nm is then spin-coated on the silicon nitride film. After that, the focusing metalens pattern is transferred onto ZEP by an exposure process with an electron-beam lithography (EBL) system at 100 kV. A 30-nm-thick Cr layer, as a hard mask, is deposited by electron beam evaporation. The residual pattern of the Cr layer on ZEP is then removed by a lift-off process combined with $O_2$ plasma cleaning. Subsequently, the 600-nm-thick silicon nitride layer is etched through by reactive ion etching (RIE). After removing the residual Cr layer on the nano bricks with the stripping solution (ceric ammonium nitrate), the designed focusing metalens with a number of regular nanobricks is obtained.

Focusing vortex metalens

A 600-nm-thick silicon nitride film is deposited on a quartz substrate by a PECVD process. Positive resist ZEP520A with a thickness of 360 nm is then spin-coated on the silicon nitride film. After that, the focusing vortex metalens pattern is transferred onto ZEP by an exposure process with an EBL system at 100 kV. A 100-nm-thick Cr layer, as a hard mask, is deposited by electron beam evaporation. The residual pattern of the Cr layer on ZEP is then removed by a lift-off process combined with $O_2$ plasma cleaning. Subsequently, the 600-nm-thick silicon nitride layer is etched through by RIE. After removing the residual Cr layer on the nano bricks with the stripping solution (ceric ammonium nitrate), the designed focusing vortex metalens with a number of regular nanobricks is obtained.

**Author Contributions**

Chuanjian Zheng and Wenli Wang: performed the calculations and numerical analyses and finished the experiment and manuscript; Yao Hu and Shaohui Zhang: supervised the experiments and finished the manuscript revision; Yanfang Ji: supervised the fabrication of the metalenses; Qun Hao: supervised the overall study.

**Data availability**

The data that support the plots within this paper and other findings of this study are available from the authors upon reasonable request. See author contributions for specific data sets.

**Conflicts of interest**

There are no conflicts to declare.

**Acknowledgments**

This work was supported by the National Key R&D Program of China (2023YFF0718101) and the National Natural Science Foundation of China (62275020).